\documentstyle[12pt,amssymb]{article}

\def\d{\partial}
\newcommand{\eq}[1]{(\ref{#1})}

\def\e{{\rm e}}

\def\d{\partial}
\def\l{\left(}
\def\r{\right)}

\def\L{{\cal{L}}}

\begin{document}
\title{%
\begin{flushright}
\normalsize UNIL-IPT-00-25\\
ULB-TH-00-27
\end{flushright}
\protect\vspace{5mm}
Three fermionic generations on a topological defect in extra
dimensions. 
}
\author{M.V.Libanov$^{1,3}$ and S.V.Troitsky$^{1,2,3}$\\
\small\em
$^{1}$~Institute for Nuclear Research of the Russian Academy of Sciences,\\
\small\em
60th October Anniversary Prospect 7a, Moscow 117312 Russia;\\
\small\em
$^{2}$~Institute of Theoretical Physics, University of Lausanne,\\
\small\em
CH-1015, Lausanne, Switzerland;\\
\small\em
$^{3}$~Service de Physique Th\'{e}orique, CP 225,\\
\small\em
  Universit\'{e} Libre de Bruxelles, B--1050, Brussels, Belgium
}
\date{}
\maketitle
\vspace{-12mm}
\begin{abstract}
We suggest a mechanism explaining the origin of three generations of
the Standard Model fermions from one generation in a
higher-dimensional theory. Four-dimensional fermions appear as
zero modes trapped in the core of a topological defect with
topological number three. We discuss hierarchical pattern of masses
and mixings which arises in these models.
\end{abstract}
\newpage
\section{Introduction}
Recently, the interest has been renewed to particle physics models in
more than four spacetime dimensions (see, for instance,
Refs.\cite{D}). These models provide an interesting framework for
solving hierarchy problems of the Standard Model of particle
interactions. In particular, it has been pointed out \cite{overlaps}
that by introducing extra dimensions it is possible to explain the
hierarchical fermionic mass pattern. The Standard Model fermions as
well as Higgs boson are represented by localized modes in extra
dimensions \cite{RubShap}, and their effective four-dimensional Yukawa
couplings are determined by overlaps between Higgs and fermionic wave
functions.  To produce hierarchical structure in overlaps, hence in
four-dimensional Yukawa couplings, it has been proposed to localize
Higgs and three fermionic generations at different points in extra
dimensions \cite{overlaps}.

In this paper, we follow a completely different approach to produce
fer\-mi\-onic spectrum via wave function overlaps, explaining
simultaneously the origin of three generations of fermions with similar
quantum numbers. Suppose one has single fermionic generation in
a multi-dimensional theory. Let us consider a topological defect
whose core corresponds to our four-dimensional world. Chiral fermionic
zero modes may be trapped in the core due to specific interaction
with the fields which build up the defect
\cite{JackiwRossi,Witten,RubShap}. In some cases, the index theorem
guarantees that the number of chiral zero modes is equal to the
topological number of the defect (see for instance Refs.\cite{index}).
We will use this property to obtain three fermionic generations localized
on a defect with topological number three while having only one
generation in the bulk. If the Higgs scalar couples to the defect, it
can also be trapped in the core \cite{Witten,RubShap}. Hierarchy
between masses of three fermionic modes arises due to their different
profiles in extra dimensions.

To be specific, we will work with the simplest topological defect
which can be characterized by an integer topological number, the
global vortex (see Sec.~2). We first recall the mechanism of
localization of fermions (Sec.~3) and scalars (Sec.~4), then explain
the origin of different Yukawa couplings for similar fermions of three
generations (Sec.~5). To explain mixing angles between fermions of
different generations, we then introduce a complication in the
model (Sec.~6). We present a general discussion of the presented class of models in Sec.~7.
Notations and technical details are outlined in Appendices.

\section{A global vortex with topological number $k$.}
Consider a theory of a complex scalar field $\Phi$ in six dimensions,
\begin{equation}
\L_\Phi=|\d_A\Phi|^2-{\lambda\over 2}\l |\Phi|^2-v^2\r^2.
\label{2*}
\end{equation}
(See Appendix \ref{notations} for notations).
The global $U(1)_g$ symmetry $\Phi\to\Phi\e^{i\alpha}$
is broken spontaneously by the vacuum expectation value $|\Phi|=v$.

Let us consider field configurations which do not depend on
$x_\mu$. We introduce polar coordinates $(r,\theta)$ in
$(x_4,x_5)$ plane. There are solutions to the classical field
equations which have the form
\begin{equation}
\Phi=v\e^{ik\theta}F(r), ~~~k=\pm 1, \pm 2, \dots.
\label{2**}
\end{equation}
Their topological numbers are defined by the winding numbers $k$; the
function $F(r)$ satisfies the following boundary conditions:
$$
F(r)\to 1, r\to\infty;
$$
$$
F(r)\to 0, r\to 0,
$$
and the ordinary differential equation,
$$
F''+{1\over r}F'-{k^2\over r^2}F-\lambda v^2 F \l F^2-1\r=0
$$
(hereafter prime denotes the derivative with respect to $r$; we assume
$k>0$).  Analytical solution to this equation is unknown. However,
from this equation, it follows that
\begin{equation}
\begin{array}{c}
F(r)=O(r^{k}),~ r\to 0;\\
F(r)=1-\frac{k^2}{2\lambda v^2r^2}+O(r^{-4}), ~ r\to\infty.
\end{array}
\label{N*}
\end{equation}

So, the field configuration describes a ``3-brane'' in six-dimensional spacetime,
located at $x_4=x_5=0$, of radial size of order
$\l\sqrt{\lambda}v\r^{-1}$. 

Energy of this configuration diverges logarithmically at large
distances, $r\to\infty$. This can be cured either by gauging $U(1)_g$
(the case of local, or Abrikosov--Nielsen--Olesen vortex) or by
introducing some cutoff at large distances (for example, putting an
``anti--vortex'' with winding number $(-k)$ far away from the
vortex). In any case, this would improve the behavior of energy
density at large distances. Here, we are interested in physics inside
the core which remains intact after these modifications.

\section{Fermionic modes.}
\label{fermions}
Consider a six-dimensional fermionic field $\Omega$ which has an axial
charge 1/2 under $U(1)_g$,
\[
\Omega\to{\rm e}^{i\frac{\alpha}{2}\Gamma_7}\Omega\;,
\]
and interacts with $\Phi$ via axial Yukawa coupling,
\begin{equation}
{\cal L}_\Omega=i\bar{\Omega}\Gamma^A\partial_A\Omega-\left(g\Phi\bar{\Omega}
\frac{1-\Gamma_7}{2}\Omega+{\rm h.c.}\right)\;.
\label{ML0}
\end{equation}
We will assume that $g$ is real: its complex phase can be rotated away
by an axial transformation of the fermion.

Let us perform Fourier transform with respect to four-dimensional coordinates
$x^\mu$,
\[
\Omega(x_\mu;x_4,x_5)=\frac{1}{(2\pi)^{3/2}}\int\!d^4k {\rm e}^{-ik_\mu x^\mu}
\Psi(k_\mu;x_4,x_5)\;,
\]
The Dirac equation following from the Lagrangian Eq.(\ref{ML0}) 
takes the form of Schr\"{o}dinger type equation for $\Psi$ in the vortex 
background (\ref{2**}),
\begin{equation}
k_0\Psi=C_ik_i\Psi+D\Psi\;,
\label{ML1}
\end{equation}
where $C^i=\Gamma^0\Gamma^i$, and
\[
D=\Gamma^0\left(-i\Gamma^4\partial_4-i\Gamma^5\partial_5+
g\Phi\frac{(1-\Gamma_7)}{2}+g\Phi^*\frac{(1+\Gamma_7)}{2}\right)\;.
\]
The operators $C_i$ and $D$ anticommute, 
$$C_iD+DC_i=0.$$
This means in
particular that one can search for a solution of Eq. (\ref{ML1}) as an
expansion in the eigenvectors $\Psi_m$ of the operator $D$,
\begin{equation}
D\Psi_m=m\Psi_m\;.
\end{equation}
There may exist a set of discrete eigenvalues $m$ with separation of
order $gv$, and continuous spectrum starting from $m\gtrsim gv$.  For
given $k_\mu$, the eigenvalues $m$ satisfy $k_0^2=k_i^2+m^2$. This
means that to excite a mode with non-zero $m$, energy of order $gv$ is
required.  In what follows we assume that the energy scales probed by
a four-dimensional observer are much smaller than $gv$, and thus even
the first non-zero level is not excited. So, we are interested only in
zero modes of $D$:
\begin{equation}
D\Psi=0\;.
\label{ML3}
\end{equation}

As is shown in Appendix \ref{fermioniczero}, 
the solution for $\Omega(x_\mu,x_4,x_5)$ has the form
\begin{equation}
\Omega(x)=\frac{1}{(2\pi)^{3/2}}\int\!d^4k \delta(k^2){\rm e}^{-ik_\mu x^\mu}
\sum_{p=0}^{k-1} a_{\bf k}^{p}
\left(
\begin{array}{c}
0\\
f_{p}(r){\rm e}^{ip\theta}c_{\bf k}\\
f_{k-p-1}(r){\rm e}^{-i(k-p-1)\theta}c_{\bf k}\\
0
\end{array}
\right)\;,
\label{NEW10*}
\end{equation}
where $c_{\bf k}$ is a normalized four-dimensional left-handed spinor,
$a_{\bf k}^p$ are arbitrary complex functions of a three-dimensional
momentum ${\bf k}=\{k_i\}$, and
$f_l(r)$ is the normalized solution to the following
differential equation,
\begin{equation}
f_l''-\l {F'\over F}+{k-1\over r} \r f_l'+ \l{F'\over F} {l\over r} +
{l(k-l)\over r^2} - g^2v^2F^2 \r f_l=0.
\label{6*}
\end{equation}
At $r\to0$, $f_l(r)\sim r^l$, while at $r\to\infty$, 
$f_l(r)\sim{\rm e}^{-gvr}$.

In other words, there exist $k$ zero fermion modes with different
angular and radial wave functions. One bulk fermion corresponds to $k$
four-dimensional fermion species.

Let us emphasize  that at given $p$, there is only
one degree of freedom for particle ($k_0>0$) and one degree of freedom 
for antiparticle ($k_0<0$)
that corresponds to one spin state of a left-handed spinor. Since
$f_p(r)$  fall off exponentially at large $r$, the particles described by
$\Omega$ are localized in the core of the vortex.

It is instructive to study how this eight-component localized spinor can be
considered from the four-dimensional point of view. For this purpose,
let us introduce a {\em $8\times8$} representation of {\em four-dimensional} 
$\gamma$-matrices: $\tilde{\gamma}^\mu=\Gamma^\mu$,
\[
\tilde{\gamma}^5=i\Gamma_0\Gamma_1\Gamma_2\Gamma_3=
\left(
\begin{array}{cc}
-\gamma^5&0\\
0&\gamma^5
\end{array}
\right)\;.
\]
These matrices together with the unit matrix, their commutators 
$\tilde{\sigma}^{\mu\nu}=
\frac{i}{2}[\tilde{\gamma}^\mu \tilde{\gamma}^\nu]$, and $\tilde{\gamma}^\mu
\tilde{\gamma}^5$ form an algebra $\tilde{\cal{A}}$ (that is any
product of these matrices is their linear combination). The algebra 
$\tilde{\cal{A}}$ is isomorphic to algebra $\cal{A}$ of $\gamma^\mu$,
and one can construct an operator which relates
$\tilde{\gamma}^\mu$ and $\gamma^\mu$:
\begin{equation}
U^+\tilde{\gamma}U=\gamma\;,
\label{MLF}
\end{equation}
for any $\tilde{\gamma}\in\tilde{\cal{A}}$, $\gamma\in\cal{A}$. 
One can  check that
\begin{equation}
U=\frac{1}{2}
\left(
\begin{array}{c}
\gamma_0(1-\gamma_5)\\
1+\gamma_5
\end{array}
\right)=
\left(
\begin{array}{cc}
0&1\\
0&0\\
1&0\\
0&0\\
\end{array}
\right)
\label{MLF1}
\end{equation}
is one of the solutions to Eq.(\ref{MLF}). To find four-component spinor
$\omega(x)$ which corresponds to $\Omega(x)$, one should act by $U^+$
on $\Omega$:
\[
\omega(x)=
U^+\Omega(x)=\frac{1}{(2\pi)^{3/2}}\int\!d^4k \delta(k^2){\rm e}^{-ik_\mu x^\mu}
\sum_{p=0}^{k-1} a_{\bf k}^{p}f_{p}(r){\rm e}^{ip\theta}
\left(
\begin{array}{c}
c_{\bf k}\\
0
\end{array}
\right)
\]
\[
\equiv
\sum_{p=0}^{k-1}f_{p}(r){\rm e}^{ip\theta}\psi_p(x_\mu)\;.
\]
We again obtain the same result: $\omega$ describes $k$ left-handed massless
spinors $\psi_p(x_\mu)$ localized at the vortex.

To localize $k$ right-handed massless spinors at the vortex, one should
consider six-dimensional spinor $\Xi$ which has an axial charge $-1/2$
under $U(1)_g$: with obvious modifications, the above analysis goes
through for $\Xi$ as well.

\section{Higgs field.}
Let us introduce a complex scalar field $H$ which interacts with the
vortex field $\Phi$:
\begin{equation}
\L_H=|\d_A H|^2-{\kappa\over 2} \l |H|^2-\mu^2\r^2-h^2|H|^2|\Phi|^2.
\label{7**}
\end{equation}
Note that at this point $H$ can have arbitrary charge under
$U(1)_g$. Later on, we will set this charge to zero. 

The system of the two scalar fields, $\Phi$ and $H$, described by the
Lagrangian $\L=\L_\Phi+\L_H$, Eqs.\eq{2*} and \eq{7**}, admits several
nontrivial classical solutions. One of them corresponds to $\Phi$
given by Eq.\eq{2**} and $H=0$. It can be shown \cite{Witten},
however, that this solution is unstable in a certain region of parameter
space, in particular, for $\kappa\mu^2\lesssim h^2v^2$. We will
consider this case in what follows; the lowest energy solution in the
topological sector where $\Phi$ has the form \eq{2**} is
\begin{equation}
\begin{array}{c}
\Phi=v\e^{ik\theta}F_c(r);\\
H=H_c(r),
\end{array}
\label{I*}
\end{equation}
where radial functions $F_c(r)$, $H_c(r)$ satisfy the following set of
nonlinear differential equations,
\begin{equation}
\begin{array}{c}
F_c''+{1\over r}F_c'-{k^2\over r^2}F_c-h^2F_cH_c^2-\lambda v^2 F_c 
\l F_c^2-1\r=0,\\
H_c''+{1\over r}H_c'-h^2 H_c v^2F^2-\kappa H_c (H_c^2-\mu^2)=0.
\end{array}
\label{7*}
\end{equation}
The boundary conditions are 
$$
F_c(0)=0,~~F_c(\infty)=1;
$$
$$
H_c(0)=v_H\ne 0,~~H_c(\infty)=0,
$$
where $v_H$ is the vacuum expectation value of the Higgs field from
four-dimensional point of view. The leading behavior of $H_c(r)$ near
the origin is
\begin{equation}
H(r)=v_H\left(1-\frac{\kappa(\mu^2-v_H^2)}{4}r^2\right)\;.
\label{higgs0}
\end{equation}
So, to satisfy to the boundary condition at $r\to \infty$, it is required 
that $v_H<\mu$. The leading behavior of $F_c(r)$ is the same as 
of $F(r)$, Eq.\eq{N*}. In the background of the modified vortex solution,
Eq.\eq{I*}, one still recovers all results of Sec.~\ref{fermions},
replacing $F(r)$ by $F_c(r)$.

\section{Four-dimensional fermion masses.}
\label{Section:fermionmasses}

Let us turn to the  fermion--Higgs Yukawa
couplings. To describe a ``prototype'' generation, we need two
six-dimensional spinors, $Q$ and $U$, with opposite charges under
$U(1)_g$:
$$
Q\to\e^{i{\alpha\over 2}\Gamma_7}Q,~~~
U\to\e^{-i{\alpha\over 2}\Gamma_7}U.
$$
The interaction with the vortex with $k=3$,
$$
g_q\Phi\bar Q{1-\Gamma_7\over 2} Q 
+g_u\Phi^*\bar U{1-\Gamma_7\over 2} U 
+{\rm h.c.}
$$
results in the existence of three left-- and right--handed zero modes of $Q$
and $U$, respectively:
\begin{equation}
\begin{array}{c}
\displaystyle
Q=
\sum\limits_{p_q=0}^{2}
\frac{1}{(2\pi)^{3/2}}\int\!d^4k\,\delta(k^2)\e^{-ik_\mu x_\mu}
a^{(q)}_{p_q}({\bf k})
\l
\begin{array}{c}
0 \\
c_{\bf k}q_{p_q}(r)\e^{ip_q\theta}\\
c_{\bf k}q_{2-p_q}(r)\e^{-i(2-p_q)\theta}\\
0 \\
\end{array}
\r
,
\\
~\\
\displaystyle
U=
\sum\limits_{p_u=0}^{2}
\frac{1}{(2\pi)^{3/2}}\int\!d^4k\,\delta(k^2)\e^{-ik_\mu x_\mu}
a^{(u)}_{p_u}({\bf k})
\l
\begin{array}{c}
d_{\bf k}u_{2-p_u}(r)\e^{-i(2-p_u)\theta} \\
0\\
0\\
d_{\bf k}u_{p_u}(r)\e^{ip_u\theta} \\
\end{array}
\r\;.
\end{array}
\label{8**}
\end{equation}
Here, $c_{\bf k}$ and $d_{\bf k}$ are  the left-handed and 
right-handed four-dimensional spinors, respectively; arbitrary complex functions
$a^{(q)}_{p_q}({\bf k})$ for each $p_q=0,1,2$ describe two degrees of
freedom (one for the particle and one for the antiparticle) of a
massless four-dimensional left-handed fermion $Q_{p_q}$ with gauge and
global quantum numbers of $Q$; $a^{(u)}_{p_u}({\bf k})$ describe a
massless four-dimensional right-handed fermion $U_{p_u}$ with quantum
numbers of $U$; $p_q=0,1,2$ corresponds to three generations of
fermions. The wave function profiles in extra dimensions are defined
by radial functions $q_{p_q}(r)$, $u_{p_u}(r)$, which are the
solutions of Eqs.\eq{6*} with $k=3$, $f=q$ or $u$, $g=g_q$ or $g_u$,
and $l=p_q$ or $p_u$, respectively.

Fermion masses originate from the following term in the
six-dimensional action:
$$
Y_u\int\!d^6x\, \tilde{H}\bar Q{1-\Gamma_7\over 2}U
%+
%Y_d\int\!d^6x\, H\bar Q{1-\Gamma_7\over 2}D+
%Y_l\int\!d^6x\, H\bar Q{1-\Gamma_7\over 2}E 
+{\rm h.c.}
$$
($Y_{u}$ is the six-dimensional Yukawa coupling constant, and
$\tilde H_i=\epsilon_{ij}H_j^*$).  This
interaction is $U(1)_g$ invariant only if $H$ is neutral under
$U(1)_g$. To obtain the effective four-dimensional mass matrix, one has to
perform the integration over extra dimensions in the action integral.
This results in the Dirac mass terms, 
$$
m_{p_qp_u}\bar Q_{p_q} U_{p_u},
$$
with
\begin{equation}
m_{p_qp_u}= Y_u\int\! r\, dr\,d\theta\,H_c(r)q_{p_q}(r) u_{p_u}(r)
\e^{i(p_q-p_u)\theta}. 
\label{8*}
\end{equation}

Integration over $\theta$ in Eq.\eq{8*} leads to a selection rule,
$m_{p_qp_u}\sim\delta_{p_q p_u}$. This means that in this way it is possible to
generate only diagonal mass terms, but not inter-generation
mixings. To obtain mixings between fermions of different generations,
it is necessary to have a non-trivial $\theta$ dependence in the Higgs
mode. We will return to this point in Sec.~6.

To see that the hierarchy of diagonal mass values can be generated in
this way, one can make use of the following, very rough,
estimation. Characteristic distance scales for the Higgs and fermionic
modes are $(\sqrt{\kappa}\mu)^{-1}\sim (hv)^{-1}$ and $(gv)^{-1}$,
respectively. Let us take $h\gg g$, so that the Higgs mode is narrow
in comparison to fermionic modes. Then one can substitute
fermionic radial functions in Eq.\eq{8*} by their leading behavior at
$r\to 0$,
$$
q_p\sim (g_qvr)^p,\ \ u_p\sim (g_uvr)^p
$$
and to write $H_c=H_c(hvr)$. In this approximation, one gets
\begin{equation}
m\sim\l {g_q g_u\over h^2} \r ^{p}\equiv\delta_u^{2p},
\label{mdelta}
\end{equation}
up to some  slowly varying function on $p$,$g$, and
$h$. Thus, at $\delta\sim 0.1$, mass hierarchy among three generations
is naturally reproduced, if we associate the first generation with
$p=2$, the second one with $p=1$ and the third one with $p=0$. To
obtain exact predictions, one has to solve the equations \eq{6*},
\eq{7*} and to evaluate the radial integral in \eq{8*} numerically. 
A similar procedure yields masses of down--quarks and charged leptons,
starting from interactions Eqs.~\eq{vff}, \eq{vhf} (see below).
Since dependence on $g$ of wave function overlaps is highly nonlinear,
the model does not predict any simple analytic relation between masses
of fermions of different kinds (for example, $m_t/m_c\ne m_b/m_s$, etc.).

The parameters of the model with all Standard Model fermions included
are: an overall mass scale, say, $Y_u \sqrt{\kappa}\mu$; two ratios of
six-dimensional Yukawa couplings, $Y_d/Y_u$ and $Y_l/Y_u$; and five
ratios $g_{q,u,d,l,e}/h$. As follows from the discussion above (see
Eq.~\eq{mdelta}), to the leading approximation, masses depend only on
three combinations of the latter five ratios, namely, $\delta_u=g_q
g_u/h^2$, $\delta_d=g_q g_d/h^2$, and $\delta_l=g_l g_e/h^2$.  Thus,
diagonal masses of nine charged fermions of the Standard Model arise
from eight independent parameters, with significant dependence from
six of them.

\section{Mixing between generations.} 

To obtain off-diagonal mass matrix elements which mix fermions of
different generations, one has to relax the selection rule
$\delta_{p_q p_u}$ in Eq.\eq{8*} by introducing non-trivial $\theta$
dependence in the Higgs mode. To do this, one has to complicate the
model, because the interaction \eq{7**} is phase--independent, so the
classical solution $H_c$ depends only on $r$. In what follows, we will
need a somewhat more complicated topological defect, a global vortex
made of two scalar fields with different winding numbers which appears
in a model with
$$
\L=\L_\Phi+\L_X,
$$
where $\L_\Phi$ is presented in Eq.\eq{2*}, and 
$$
\L_X=|\d_A X|^2-{\lambda_1\over 2}\l |X|^2-v_1^2 \r^2-\alpha \l
X^3\Phi^*+
X^{*3}\Phi \r.
$$
Complex six-dimensional scalar fields $\Phi$ and $X$ have charges 3
and 1 under global $U(1)_g$ symmetry, respectively\footnote{We assume
that $\alpha$ is real; this can always be reached by rotating away its
complex phase via redefinition of the origin of polar angle
$\theta$.}. Let $\tilde v$ and $\tilde v_1$ be vacuum expectation
values of $|\Phi|$ and $|X|$,
respectively; 
$$
\tilde v=v +O(\alpha),
$$
$$
\tilde v_1=v_1 +O(\alpha).
$$
At $\alpha<\lambda_1$, the model admits stable global vortices of the form
\begin{equation}
\begin{array}{rcl}
\Phi=\tilde v \e^{3i\theta}\tilde F(r),\\
X=\tilde v_1 \e^{i\theta} \chi(r),
\end{array}
\label{10*}
\end{equation}
with $\tilde F(r)$ different than $F(r)$ in Eq.\eq{2**}, but still
having the same leading behavior at $r\to 0$ and $r\to\infty$.  Note
that in this model, the solution, Eq.~(\ref{10*}), is the simplest
nontrivial vortex.

Let us couple six-dimensional fermions corresponding to five fermionic fields of one
generation of the Standard Model (namely, left-handed leptons $L$ and quarks $Q$
and right-handed leptons $E$, up quarks $U$ and down quarks $D$) to
$\Phi$ in a way similar to Eq.\eq{ML0},
\begin{equation}
\begin{array}{rcl}
V_{\Phi f}&=&
g_q\Phi\bar Q_i{1-\Gamma_7\over 2} Q_i +
g_l\Phi\bar L_i{1-\Gamma_7\over 2} L_i +\\
&&g_u\Phi^*\bar U{1-\Gamma_7\over 2} U +
g_d\Phi^*\bar D{1-\Gamma_7\over 2} D +
g_e\Phi^*\bar E{1-\Gamma_7\over 2} E +
{\rm h.c.} ,
\end{array}
\label{vff}
\end{equation}
where we have written explicitly the electroweak $SU(2)$ indices $i$,
and to the Higgs field $H$,
\begin{equation}
V_{Hf}=Y_u \tilde H_i\bar Q_i{1-\Gamma_7\over 2} U+
Y_d H_i\bar Q_i{1-\Gamma_7\over 2} D +
Y_l H_i\bar L_i{1-\Gamma_7\over 2} E +
{\rm h.c.},
\label{vhf}
\end{equation}
where $\tilde H_i=\epsilon_{ij}H_j$. 
Three chiral generations are localized in four dimensions
as zero modes of the form \eq{8**} (with $F(r)$ replaced by
$\tilde F(r)$ in corresponding equations). 

On the other hand, let the 
Higgs field $H$ couple to the field $X$ in a way similar
to Eq.\eq{7**},
\begin{equation}
V_{\Phi H}={\kappa\over 2}\l |H_i|^2-\mu^2\r ^2 +h^2 |H_i|^2 |X|^2 +\Delta V,
\label{11*}
\end{equation}
where $\Delta V$ is a small perturbation which breaks $U(1)_g$ global
symmetry explicitly, for instance,
$$
\Delta V= |H_i|^2(\epsilon X +\epsilon^* X^* ).
$$
At $\epsilon=0$, we recover Eq.\eq{7*} with $F_c(r)$ replaced by
$\chi(r)$; its classical solution, call it $H_0(r)$, contributes to
the diagonal masses through the integral Eq.\eq{8*}. If $\epsilon\ne
0$, the solution to the nonlinear partial differential equation for
$H(r,\theta)$ which follows from \eq{11*} in the background \eq{10*}
has a complicated $\theta$ dependence. For our purposes, however, it
is sufficient to perform perturbative expansion,
\begin{equation}
H(r,\theta)=H_0(r)+\epsilon H_1(r,\theta) +\epsilon^2
H_2(r,\theta)+\dots+{\rm h.c.}
\label{exphiggs}
\end{equation}
The first term is $H_0(r)$ defined above. In the first order in $\epsilon$, a
source term of the form
$$
\tilde v_1 \chi(r) \e^{i\theta} H_0(r),
$$
appears in the equation for $H_1(r,\theta)$ which results in 
$$
H_1(r,\theta)=\e^{i\theta}h_1(r).
$$
In a similar way,
$$
H_2(r,\theta)=\e^{2i\theta}h_2(r).
$$
The integral \eq{8*} reads now\footnote{Hereafter, the indices
$a,b=1,2,3$ are used instead of $p_q,p_u,\dots=2,1,0$ to enumerate the
fermionic generations; $a=1$ corresponds to the first generation, etc.}
\begin{equation}
\begin{array}{rl}
m_{a b}^u&=
Y_u\int\! r\, dr\,d\theta\,H^*(r,\theta)q_{3-a}(r) u_{3-b}(r)
\e^{i(b-a)\theta}= \\
&
\begin{array}{rl}
=Y_u(2\pi)\int\! r\, dr\,q_{3-a}(r) u_{3-b}(r)&\l
H_0(r)\delta_{ab}+\right.\\
&\epsilon h_1(r) \delta_{a-1,b}+\epsilon^* h_1(r) \delta_{a,b-1}+\\
&\left. \epsilon^2 h_2(r) \delta_{a-2,b}+\epsilon^{*2} h_2(r)
\delta_{a,b-2}+
\dots \r.
\end{array}
\end{array}
\label{massmatrix}
\end{equation}
Non-diagonal mass matrix elements arise from overlaps of $h_{1,2}(r)$
with fermionic wave functions and are suppressed by powers of
$\epsilon$ with respect to diagonal ones. 

In the form discussed above, the model does not admit $CP$ violation
since the overall phase of $Y_u$ is irrelevant and phase of $\epsilon$ 
can be included in
two-component spinors $c_{\bf k}$ and $d_{\bf k}$ (see
Eq.(\ref{8**})).  With more complicated $\Delta V$, for example,
\begin{equation}
\Delta V=\epsilon X |H_i|^2+\epsilon_1 X^2 |H_i|^2 +{\rm h.c.},
\label{dletav}
\end{equation}
the relative phase of $\epsilon_1$ and $\epsilon$ is a free $CP$
violating parameter of the model (while all other results remain
intact).

The relevant real parameters are now $g_{q,u,d,l,e} \tilde{v}/(h\tilde
v_1)$, $|\epsilon|$, $|\epsilon_1|$, $\arg(\epsilon_1-\epsilon)$ and
again $Y_u\sqrt{\kappa}\mu$ and $Y_{d,l}/Y_u$. These eleven parameters
of the six-dimensional theory thus generate thirteen parameters of the
fermionic sector of the Standard Model (nine diagonal masses, three
mixing angles, and one CP violating phase). As before, results depend
significantly not on all five ratios $g\tilde{v}/(h\tilde v_1)$, but
only on three combinations $g_q g_u\tilde{v}^2/(h^2\tilde{v_1}^2)$,
$g_q g_d\tilde{v}^2/(h^2\tilde{v_1}^2)$, 
$g_l g_e\tilde{v}^2/(h^2\tilde{v_1}^2)$.

To demonstrate that it is possible to obtain a realistic hierarchical
pattern of the fourteen Standard Model masses and mixings, let us
estimate them in more detail. First of all, we find
$H(r,\theta)$ (see Eq.(\ref{exphiggs})) in the model with $\Delta V$
given by Eq.(\ref{dletav}). We assume that
$\epsilon$ and $\epsilon_1$ are of the same order and find
$H(r,\theta)$ in the first order in $\epsilon$, $\epsilon_1$:
\begin{equation}
H(r,\theta)=H_0(r)+h_1(r)(\epsilon\e^{i\theta}+{\rm h.c.})+
h_2(r)(\epsilon_1\e^{2i\theta}+{\rm h.c.}).
\label{higgsee1}
\end{equation}
Close to the origin 
\begin{equation}
H_0(r)= v_H;~~~ 
h_1(r)= \frac{v_HX'(0)}{8}r^3;~~~ h_2(r)=\frac{v_HX'(0)^2}{12}r^4,
\label{behavior}
\end{equation}
(recall that $X$ has winding number one and thus 
$X'(0)\neq 0$). On the other hand, the width of $h_{1,2}$ is
of order of width of $H_0$ that is of order $(hv)^{-1}$. Substituting
Eq.(\ref{behavior}) into Eq.(\ref{massmatrix}) and estimating all
integrals as it has been done it in Sec.~\ref{Section:fermionmasses}, 
we obtain the following mass matrix of up--quarks
\[
m_{ab}^u
\sim Y_u
\left(
\begin{array}{ccc}
\delta^2_u&\epsilon^*\delta^3_u&\epsilon_1^*\delta^3_u\\
\epsilon\delta^3_u&\delta_u&\epsilon^*\delta^2_u\\
\epsilon_1\delta^3_u&\epsilon\delta^2_u&1\\
\end{array}
\right)
\]
and a similar matrix with the replacement $\delta_u\to \delta_d$, $Y_u\to
Y_d$, for down quarks. 

Cabibbo-Kobayashi-Maskawa (CKM) mixing
matrix is defined by
$$
U^{CKM}=S^\dagger_u S_d,
$$
where $S_u$ and $S_d$ transform the matrices $m_u m_u^\dagger$ and 
$m_d m_d^\dagger$, respectively, to the diagonal form:
$$
S^\dagger m m^\dagger S=\mbox{diag}(\dots).
$$

With the definitions
\[
A_u=\left|\frac{m_{21}^um_{11}^u+m_{12}^um_{22}^u}{m_{22}^{u2}-m_{11}^{u2}}
\right|\sim |\epsilon|\delta^2\;;\
\
B_u=\left|\frac{m_{23}^um_{33}^u+m_{32}^um_{22}^u}{m_{33}^{u2}-m_{22}^{u2}}
\right|\sim |\epsilon|\delta^2\;;
\]
\[
C_u=\left|\frac{m_{13}^um_{33}^u+m_{31}^um_{11}^u}{m_{33}^{u2}-m_{11}^{u2}}
\right|\sim |\epsilon_1|\delta^3\;;\ \
\]
and the same for down quarks we find that in the leading order,
\begin{eqnarray}
U^{CKM}\simeq&\left(
\begin{array}{ccc}
1&(A_d-A_u)&(C_d-C_u)\\
(A_u-A_d)&1&(B_d-B_u)\e^{-i\varphi}\\
(C_u-C_d)&(B_u-B_d)\e^{i\varphi}&1\\
\end{array}
\right)
\nonumber\\
&\sim\left(
\begin{array}{ccc}
1&|\epsilon|(\delta_d^2-\delta_u^2)&|\epsilon_1|(\delta_d^3-\delta_u^3)\\
|\epsilon|(\delta_u^2-\delta_d^2)&1&|\epsilon|(\delta_d^2-\delta_u^2)\e^{-i\varphi}\\
|\epsilon_1|(\delta_u^3-\delta_d^3)&|\epsilon|(\delta_u^2-\delta_d^2)\e^{i\varphi}&1\\
\end{array}
\right)\nonumber
\end{eqnarray}
where $\varphi=\arg{\epsilon_1}-2\arg{\epsilon}$.
We see that there is a hierarchy in CKM
matrix in our model which coincides with the hierarchy of CKM in
Standard Model: the magnitude of the matrix elements decreases 
away from the diagonal, so that $V_{cb}\sim
V_{us}\sim\epsilon\delta^2$, $V_{ub}\sim\epsilon_1\delta^3$.

\section{Conclusions.}
The mechanism discussed here provides a consistent picture which can
explain the origin of three generations of fermions with identical
gauge and global quantum numbers but hierarchical mass matrices
without  fine tuning of parameters. Chiral fermionic zero
modes are localized on a topological defect with topological number
three which explains the origin of three generations; different masses
appear due to different profiles of three fermionic zero modes in
extra dimensions. 

To explain mixings between generations, one has to introduce small but
explicit violation of the global $U(1)_g$ symmetry of the
theory.
Off--diagonal mass matrix elements are suppressed by powers of
the small parameter $\epsilon$ characterizing this violation. Though
topological arguments are no longer valid in the presence of this
$U(1)_g$ violation, we expect our results to be stable with respect to
$\epsilon$.

The model with a global vortex can be embedded either in a theory of
large compact extra dimensions (in which case the problem of stability
of the vortex in a compact space has to be addressed) or in a model
with localized gravity \cite{vortexRS}. The mechanism suggested here
works as well in models with other topological defects with
topological number three, for example, with a ``hedgehog'' in seven
space-time dimensions.

The models which exploit the mechanism discussed above are in
principle fairly 
predictive. For instance, in our particular toy model with a global
vortex, eleven independent parameters of the six-dimensional theory
generate thirteen masses and mixings of the Standard Model fermions,
and the hierarchical structure of mass matrices is reproduced.

We are indebted to V.Rubakov and M.Voloshin for numerous helpful
discussions, and to M.Schmaltz for an interesting discussion both of
the results presented here and of his unpublished work with D.Kaplan and
N.Arkani--Hamed where localization of three chiral fermionic
generations on a vortex was studied in the framework of a different
model. We appreciate stimulating conversations about localization of
chiral fermions with J.-M.Fr\`{e}re and his colleagues from
Universit\'{e} Libre de Bruxelles (ULB). We thank H.-C.Cheng,
S.Dubovsky, D.Gorbunov, D.E.Kaplan, A.Masiero, and T.Tait for
interesting discussions. S.T.\ thanks Aspen Center for Physics for
hospitality during the workshop on ``New physics at the weak scale and
beyond'' where part of this work was done.
This work was completed during the authors' stay at
ULB, which we thank for kind hospitality and for partial support by
the ``Actions de Recherche Concret\'ees'' of ``Communaut\'e
Fran\c{c}aise de Belgique'' and IISN--Belgium.  This work is supported
in part by RFFI grant 99-02-18410a, by CRDF award RP1-2103, by the
Russian Academy of Sciences, JRP grant No.~37, and by the programme
SCOPES 
of the Swiss National Science Foundation, project No. 7SUPJ062239, 
financed by Federal Department of Foreign affairs.  The work of S.T.\
is supported in part by Swiss Science Foundation, grant 21-58947.99.

\appendix
\section{Notations.}
\label{notations}

Six-dimensional coordinates $x_A$ are labeled by capital Latin indices
$A,B=0,\dots,5$. Four-dimensional coordinates $x_\mu$ are labeled by
Greek indices $\mu,\nu=0,\dots,3$; for spatial coordinates we use
lower case Latin indices $i,j=1,2,3$. The Minkowski metric is $g_{AB}={\rm
diag}
(+,-,\dots,-)$.

Dirac fermions in six dimensions are described by eight-component spinors;
we work with the following representation of six-dimensional 8$\times8$
Dirac matrices $\Gamma^A$:
\[
\Gamma^A=
\left(
\begin{array}{cc}
0&\Sigma^A\\
\bar{\Sigma}^A&0
\end{array}
\right)\;,
\]
where $\Sigma^0=\bar{\Sigma}^0=\gamma^0\gamma^0$; $
\Sigma^i=-\bar{\Sigma}^i=\gamma^0\gamma^i$; $\Sigma^4=-
\bar{\Sigma}^4=i\gamma^0\gamma^5$,
$\Sigma^5=-\bar{\Sigma}^5=\gamma^0$, and $\gamma^\mu$,
$\gamma^5$ are the usual four-dimensional Dirac
matrices in the chiral representation:
\[
\gamma^0=
\left(
\begin{array}{cc}
0&1\\
1&0
\end{array}
\right)\;,
\ \ 
\gamma^i=
\left(
\begin{array}{cc}
0&\sigma^i\\
-\sigma^i&0
\end{array}
\right)\;,
\ \
\gamma^5=i\gamma_0\gamma_1\gamma_2\gamma_3=
\left(
\begin{array}{cc}
1&0\\
0&-1
\end{array}
\right)\;,
\]
$\sigma^i$ are the Pauli matrices.

We also introduce $\Gamma_7$
which is an analog of four-dimensional matrix $\gamma_5$
\[
\Gamma_7=\Gamma_0\dots\Gamma_5=
\left(
\begin{array}{cc}
1&0\\
0&-1
\end{array}
\right)\;.
\]

\section{Fermionic zero modes and their asymptotics.}
\label{fermioniczero}

The solution to Eq.(\ref{ML3}) has the
following structure
\begin{equation}
\Psi(k_\mu;r,\theta)=
\left(
\begin{array}{l}
f_{(1)}(r)c_{(1)}{\rm e}^{i(p+1)\theta}\\
f_{(2)}(r)c_{(2)}{\rm e}^{ip\theta}\\
f_{(3)}(r)c_{(3)}{\rm e}^{-i(k-1-p)\theta}\\
f_{(4)}(r)c_{(4)}{\rm e}^{-i(k-p)\theta}
\end{array}
\right)
\label{ML4}
\end{equation}
In the last equation, $p$  is an integer number; $c_{(a)}$ 
are two component columns (which, as will be shown below,
correspond to four-dimensional
chiral spinors), and $f_{(a)}$ satisfy the following set of 
differential equations
\footnote{These equations cannot be solved analytically, except for a
particular case of $p=(k-1)/2$, when a normalized solution has the
form
$$
f_{(2),(3)}={\rm const}\cdot r^{k-1\over 2}\exp{\left( -\int\limits^r gv
F(x)dx\right)}
\;, \ \ f_{(1),(4)}=0\;.
$$
},
\begin{eqnarray}
&&\displaystyle \left\{
\displaystyle \begin{array}{l} 
\displaystyle f'_{(1)}+\frac{(p+1)}{r}f_{(1)}-gvFf_{(4)}=0,\\
\\
\displaystyle f'_{(4)}+\frac{(k-p)}{r}f_{(4)}-gvFf_{(1)}=0;
\end{array}
\right.
\label{ML514}\\
\nonumber\\
&&\displaystyle \left\{
\displaystyle \begin{array}{l}
\displaystyle f'_{(2)}-\frac{p}{r}f_{(2)}+gvFf_{(3)}=0,\\
\\
\displaystyle f'_{(3)}-\frac{(k-p-1)}{r}f_{(3)}+gvFf_{(2)}=0.
\end{array}
\right.
\label{ML523}
\end{eqnarray}

To investigate the behavior of $f_{(a)}$ and to find their asymptotics, it is 
convenient to introduce the new set of functions $\tilde{f}_{(a)}$:
\begin{equation}
f_{(1)}=r^{-(p+1)}\tilde{f}_{(1)}\;,\ \ 
f_{(2)}=r^{p}\tilde{f}_{(2)}\;,\ \ 
f_{(3)}=r^{(k-1-p)}\tilde{f}_{(3)}\;,\ \ 
f_{(4)}=r^{-(k-p)}\tilde{f}_{(4)}\;.
\label{ML6}
\end{equation}
The functions $\tilde{f}_{(a)}$ satisfy the following
set of  differential equations obtained from Eqs.(\ref{ML514}), (\ref{ML523}):
\begin{eqnarray}
\left\{
\begin{array}{l}
\displaystyle
\tilde{f}''_{(4)}-\left(\frac{F'}{F}+\frac{(k-2p-1)}{r}\right)
\tilde{f}'_{(4)}-
g^2v^2F^2\tilde{f}_{(4)}=0,\\
\\
\displaystyle \tilde{f}_{(1)}=\frac{\tilde{f}_{(4)}'}{gvFr^{k-2p-1}};
\end{array}
\right.
\label{ML714}
\\
\left\{
\begin{array}{l}
\displaystyle \tilde{f}''_{(2)}-\left(\frac{F'}{F}+
\frac{(k-2p-1)}{r}\right)\tilde{f}'_{(2)}-
g^2v^2F^2\tilde{f}_{(2)}=0,\\
\\
\displaystyle \tilde{f}_{(3)}=-\frac{\tilde{f}_{(2)}'}{gvFr^{k-2p-1}}.
\end{array}
\right.
\label{ML723}
\end{eqnarray}

Note that $\tilde{f}_{(1)}$ and $\tilde{f}_{(3)}$ satisfy the same 
differential equations as $\tilde{f}_{(4)}$ and $\tilde{f}_{(2)}$ 
respectively, with $p$ replaced by
$k-p-1$. 

>From Eqs.(\ref{ML714}) it follows that, if at some point $r_0$ a
solution $\tilde{f}_{(4)}$ and its derivative are positive
($\tilde{f}_{(4)}(r_0)\ge 0$, $\tilde{f}'_{(4)}(r_0)>0$), then
$\tilde{f}_{(4)}$ increases with $r$ at any point $r>r_0$. To see
this, let us first note that equation on $f_{(a)}$ is nothing but
a Schr\"{o}dinger type equation for the lowest energy level. This means
that $f_{(a)}$ cannot have nodes at $0<r<\infty$, so we can
assume  $f_{(4)}\ge0$. Let us assume also that $(k-1)/2 \ge p$,
and thus $k>p$.  Since $F'/F\ge 0$ and $\tilde{f}'_{(4)}(r_0)>0$, it
follows from Eq.(\ref{ML714}) that $\tilde{f}''_{(4)}(r_0)>0$. It
means that $\tilde{f}'_{(4)}$ increases and so $\tilde{f}'_{(4)}(r)>0$
at any point $r>r_0$.

Let us now study the behavior of $\tilde{f}_{(a)}$ at  $r\to 0$. In
this limit,
$F\simeq r^k$ and from Eqs. (\ref{ML714}) and
(\ref{ML723}) it follows that
\[
\tilde{f}_{(4)}\simeq r^0\ \ \mbox{or}\ \ \tilde{f}_{(4)}\simeq r^{2(k-p)}\;;\ \ 
\tilde{f}_{(1)}\simeq r^0\;,
\]
\[
\tilde{f}_{(2)}\simeq r^0\ \ \mbox{or}\ \ \tilde{f}_{(2)}\simeq r^{2(k-p)}\;;\ \
\tilde{f}_{(3)}\simeq r^0\;.
\]
Since $k> p$, the solutions which have the behavior $\tilde{f}_{(2),(4)}
\simeq r^{2(k-p)}$ increase everywhere.

Now let us consider asymptotics of $\tilde{f}_{(a)}$ at $r\to \infty$, when
$F\sim 1$ and first equation in (\ref{ML714}) reads 
\[
\tilde{f}''_{(4)}-\frac{k-2p-1}{r}\tilde{f}'_{(4)}-
g^2v^2\tilde{f}_{(4)}=0\;.
\]
This equation has two linearly independent solutions,
\[
\tilde{f}_{(4)}^{I}=r^{(k-2p)/2}K_{(k-2p)/2}(gvr)\;;\ \
\tilde{f}_{(4)}^{II}=r^{(k-2p)/2}I_{(k-2p)/2}(gvr)\;,
\]
where $I_\nu(r)$ and $K_\nu(r)$ are modified Bessel functions of the
first and the third order. The second solution $\tilde{f}_{(4)}^{II}$
grows exponentially at infinity.  On the other hand, as we have
shown above, if a solution to equations (\ref{ML714}), 
(\ref{ML723}) 
increases at some $r$, then it increases 
everywhere. This means that the solution which
behaves as $\tilde{f}_{(4)}\simeq r^{2(k-p)}$ at $r\to 0$ has the
asymptotic behavior $\tilde{f}_{(4)}^{II}$ at $r\to \infty$. Thus,
the corresponding function is not normalizable.  Therefore, all
functions $\tilde{f}_{(a)}$ which can correspond to normalizable
solutions have the same asymptotics: they tend to constant at $r=0$
and exponentially fall off at infinity.  This results in the following
behavior of functions $f_{(a)}$: they exponentially fall off at infinity,
\begin{equation}
f_{(a)}\propto {\rm e}^{-gvr}\;, \ \ r\to\infty\;,
\label{expon}
\end{equation} 
and close to  the origin,
\begin{equation}
f_{(1)}\simeq r^{-(p+1)}\;,\ \ 
f_{(4)}\simeq r^{-(k-p)}\;,\ \ 
f_{(2)}\simeq r^{p}\;,\ \ 
f_{(3)}\simeq r^{k-p-1}\;,\ \ r\to0\;.
\label{MLD}
\end{equation}
Since $k-p>0$, $f_{(4)}$ is not normalizable: the corresponding
integral diverges at $r=0$. So, one is forced to conclude that
$f_{(1)}=f_{(4)}=0$.  Moreover, to allow $f_{(2)}$ to be normalizable
one should require that $p\ge0$.

The case $ 2p\ge k-1$ can be studied in the same way: instead of functions
$\tilde{f}_{(4)}$ and $\tilde{f}_{(2)}$, one should consider functions 
$\tilde{f}_{(1)}$ 
and $\tilde{f}_{(3)}$. As a result one finds that $k-1\ge p$.
There are $k$ linearly independent normalizable
solutions which can be labeled by index $p=0,\dots,k-1$:
\[
\Psi_p=
\left(
\begin{array}{c}
0\\
f_p(r){\rm e}^{ip\theta}c_{p}\\
f_{k-1-p}(r){\rm e}^{-i(k-p-1)\theta}c_{k-p-1}\\
0
\end{array}
\right)\;,
\]
where $f_p$ are the normalized solutions of the equations (\ref{ML523}) which 
behave as $r^p$ at zero and fall exponentially at infinity.

Substituting  spinor $\Psi_{p}$ into equation
\[
k_0\Psi_{p}=C_ik_i\Psi_{p}\;,
\]
one finds that $c_{p}$ and $c_{k-p}$  satisfy the following equation
\[
(k_0+k_i\sigma_i)c_{p}=(k_0+k_i\sigma_i)c_{k-p}=0\;,
\]
which defines a left-handed four-dimensional spinor for $k_0>0$. This
equation has one solution (iff $k_0^2=k_i^2$) which we will denote as
$c_{\bf k}$.  To conclude, we have precisely $k$ solutions which describe
$k$ massless left-handed four-dimensional fermions in full agreement
with the index theorem, so the solution to Eq.(\ref{ML1}), which
corresponds to the zero mode, has the form (\ref{NEW10*}).

\end{document}